\documentclass[12pt,a4paper,showkeys]{article}
\usepackage[font=footnotesize,labelfont=bf,width=1.1\textwidth]{caption}
\usepackage{authblk}
\usepackage{epsfig,amsmath,amsfonts,amssymb,cancel}
\usepackage{graphicx,latexsym,bbold,mathbbol,color}
\usepackage{stackrel,epstopdf,xcolor,hyperref,cite}
\usepackage{slashed}
\usepackage[hang,flushmargin]{footmisc}
\usepackage{soul}
\allowdisplaybreaks

\begin{document}
\numberwithin{equation}{section}

\title{Dimensional Regularization for the Particle Transition Amplitude in Curved Space} 
      
\author[a,b]{Olindo Corradini\,\footnote{E-mail: olindo.corradini@unimore.it}}
\author[a]{Luigi Crispo\,\footnote{E-mail: 216238@studenti.unimore.it}}
\author[a,b]{Maurizio Muratori\,\footnote{E-mail: maurizio.muratori@unimore.it}}

\affil[a]{{\small\it Dipartimento di Scienze Fisiche, Informatiche e Matematiche, \protect\\[-1.5mm]
Universit\`a degli Studi di Modena e Reggio Emilia, 
Via Campi 213/A, I-41125 Modena, Italy}}

\affil[b] {{\small\it INFN, Sezione di Bologna,  Via Irnerio 46, I-40126 Bologna, Italy}}

\date{\empty}

\maketitle

\abstract{We compute the perturbative short-time expansion for the transition amplitude of a particle in curved space time, by employing Dimensional Regularization (DR) to treat the divergences which occur in some Feynman diagrams. The present work generalizes known results where DR was applied to the computation of one-loop effective actions, which in the worldline approach are linked to particle path integrals on the circle, i.e. with periodic boundary conditions. The main motivation of the present work comes from revived interest in particle transition amplitudes in curved space-times, due to their use in the recently proposed worldline quantum field theory (in curved space-time).  \\[1cm]}

\noindent

\newpage

\tableofcontents

\section{Introduction}
\label{sec:Introdution}

In flat space-time, first-quantized approaches to the computation of Quantum Field Theory observables have been known since the fifties, after a pair of renowned Feynman's papers~\cite{Feynman, Feynman2}, which devised a wordline approach to the scalar QED propagator. However, these approaches have been systematically used only after the work of Bern and Kosower~\cite{Bern}, and Strassler~\cite{Strassler}, who derived  compact master formulas for photon and gluon amplitudes at one loop. Soon after, a generalization of such formulas for the case of a scalar particle line coupled to photons was proposed by Daikouji et al~\cite{Daikouji}. For the case of worldline path integrals on the circle, i.e. with periodic boundary conditions, several applications soon followed these papers, ranging from multiloop calculations~\cite{Schmidt:1994zj}, finite temperature computations~\cite{McKeon:1992if, Venugopalan:2001hp}, the inclusion of constant external electromagnetic fields~\cite{Shaisultanov:1995tm, Adler:1996cja, Reuter:1996zm} and Heisenberg-Euler lagrangians~\cite{Dunne:2002qf}, and many more. Worldline path integrals on the line were used instead to compute, for instance, thermal Green's functions~\cite{McKeon:1993np}, to represent scalar and Dirac propagators~\cite{Casalbuoni:1974pj, Casalbuoni:1980}, as well as open line spin factors~\cite{Karanikas:1999hz}, and more recently in the worldgraph approach~\cite{Dai:2006vj, Holzler:2007xt} and to obtain master formulas from the photon-dressed Dirac propagators~\cite{Ahmadiniaz:2020wlm}---see the review papers~\cite{Schubert, Edwards-Schubert} for an extensive bibliography and a summary of classic and more recent applications.    

The inclusion of gravity in the worldline formalism was instead considered more recently, even though a generalization of the Bern-Kosower rules which includes gravity was already proposed by Bern, Dunbar and Shimada in Ref.~\cite{Bern:1993wt}. The main reason for this delay is that the use of the worldline formalism in curved spaces is linked to the construction and the computation of particle path integrals whose actions are non-linear sigma models. The treatment of such theories had been the source of some controversy in the past. However, all the fog was dispelled a couple of decades ago, and various applications have been considered ever since, ranging from one-loop effective actions and graviton amplitudes~\cite{Bastianelli:2002fv, Bastianelli:2005vk, Bastianelli:2005uy}, higher spin fields~\cite{Bastianelli:2008nm, Bastianelli:2012bn}, gravitational corrections to Heisenberg-Euler lagrangians~\cite{Bastianelli:2008cu} and spinning particle representations for Einstein gravity~\cite{Bonezzi:2018box}---see the book~\cite{Bastianelli-PvN} for a comprehensive review. However, a master formula for the computation of tree-level and loop amplitudes involving an arbitrary number of gravitons, akin to the Bern-Kosower one, is still lacking and would be quite welcome. In fact, recently there was a revival of interest on the applications of the gravitational worldline formalism. Important applications which may benefit of such formulas, are the use of worldline particle path integrals in curved space on the line, in what was recently dubbed ``worldline quantum field theory"~\cite{Mogull:2020sak}\footnote{However, in the computation of observables associated to the classical scattering of black holes, the issues associated to the regularization scheme, discussed in the present manuscript, appear not be relevant~\cite{Jakobsen:2021lvp}}, the worldline approach to the computations of amplitudes with gravitons (see e.g.~\cite{Ahmadiniaz:2019ppj}), as well as the applications of color-kinematics duality to gravity~\cite{Garozzo:2018uzj, Bern:2019prr} within the Bern-Dunbar-Shimada formalism~\cite{Bern:1993wt}, along the lines of what recently done for gluon amplitudes within the Bern-Kosower formalism~\cite{Ahmadiniaz:2021fey}.  

From the worldline viewpoint, one of the main issues to tackle is the necessity of employing a regularization scheme to correctly compute the path integrals. In fact, at the perturbative level, the presence of derivative interactions in the non-linear sigma model implies that single Feynamn diagrams can be superficially divergent or, at best, ambiguous due to the presence of products of distributions. In turn, along with the regularization, a counterterm potential is obviously needed. However, being the field theory a one-dimensional representation of particle quantum mechanics, it is overall finite and no infinite renormalization is thus necessary; i.e. the counterterm potential is finite and local. However, different regularization schemes come along with different potentials. The most used regularizations in such context are ``Time Slicing'' (TS), ``Mode Regularization'' (MR) and ``Dimensional Regularization'' (DR)---see Ref.~\cite{Bastianelli-PvN}. The main common feature of the first two is the presence of a non-covariant counterterm potential, whereas the last one only needs a covariant potential~\cite{Bastianelli:2000nm}---see also Refs.~\cite{Kleinert:1999aq, Kleinert:1999ei, Bastianelli:2000pt} for earlier constructions, and Ref.~\cite{Bastianelli:2000dw} for a higher-loop test, based on the computation of the trace anomaly for a scalar field in six dimensions. Hence, in the optics of using such theories in the computation of amplitudes with gravitons, and for the recent applications mentioned above it is helpful to single out the most efficient scheme. On the one hand, TS is the most straightforward, as it is rooted directly to the first-principled formulation of the path integral, which originates from the multiple slicing of the (operatorial) particle transition amplitude. As such, it is also the natural one to be used in non-perturbative numeric (Monte Carlo) approaches~\cite{Corradini:2020tgk}. Moreover, in perturbative computations it requires no integration by parts (i.b.p.) to compute the particle transition amplitude, yet it needs a non-covariant counterterm which may result somewhat inconvenient. On the other hand, DR does involve only a covariant counterterm but requires some specific manipulations, namely specific i.b.p.'s, to cast the worldline integrand in a form which can be computed unambiguously, removing the regulator. Moreover, DR was proposed and tested in the computation of QFT one-loop effective actions, i.e. with periodic boundary conditions. Thus, in order to make contact with the aforementioned recent work which involve particle path integrals with open boundary condition, i.e. on the line, we consider it helpful to extend and test the regularization in such context. This is the main goal of the present manuscript, and we achieve it by computing, via dimensional regularization, the short-time transition amplitude for a particle on curved space, at three loops, using Riemann Normal Coordinates, which yields the heat kernel expansion of the particle at quadratic order in the propagating time.   

The manuscript is organized as follows: in Section~\ref{sec:DR} we introduce the worldline path integrals in curved space, in Section~\ref{sec:DRTA} we describe the short-time transition amplitude and, specifically, the use of Dimensional Regularization, and we give some Conclusions in Section~\ref{sec:concl}---a pair of technical Appendices are included at the end. 

\mbox{}\\

%


\section{Worldline Path Integrals in Curved Space}
\label{sec:DR}
The classical dynamics of a unit-mass non-relativistic point particle in a $D$-dimensional curved space, and subject to a scalar potential, can be described by means of a non-linear sigma model
\begin{align}
    S[x] = \int_0^t dt~ \Big[ \frac12 g_{\mu\nu} (x) \dot x^\mu \dot x^\nu -V(x)\Big]~.
    \label{eq:Msigma}
 \end{align}
However, such action can as well be used in the worldline description of a relativistic point-particle in a $D$-dimensional curved space-time, via the Brink-Di Vecchia-Howe formulation. For this reason we prefer to use greek indices to represent the particle coordinates, which could be purely spatial for a nonrelativistic particle, or euclideanized space-time ones for a relativistic particle.  Moreover, note that in the worldline description of a relativistic scalar particle in curved space, the potential $V$ may represent the (non-minimal) coupling of the scalar field to the curvature, i.e. $V=\alpha R$.

At the quantum level, the transition amplitude for the particle described by Eq~\eqref{eq:Msigma}, satisfies the associated Schr\"odinger equation. In the present manuscript we prefer to Wick rotate the time variable, $it =:\beta$, and study the heat kernel which is the solution of the heat equation
\begin{align}
    -\partial_\beta K(x,x';\beta) = \Big[ -\frac{1}{2} \nabla^2 +V(x)\Big]  K(x,x';\beta)~,
    \label{eq:HE}
\end{align}
where $\nabla^2$ is the covariant Laplacian. The previous equation is thus linked to the Euclidean action
\begin{align}
    S[x] = \int_0^\beta dt~ \Big[ \frac12 g_{\mu\nu} (x) \dot x^\mu \dot x^\nu +V(x)\Big]~.
 \end{align}
At short `time' $\beta$, the solution to the heat equation~\eqref{eq:HE} can be expressed as the so-called heat-kernel expansion and reads
\begin{align}
  K(x,x';\beta) =\frac{1}{(2\pi \beta)^{\tfrac D2}} e^{-\sigma(x,x')/\beta} \sum_{n=0}^\infty a_n(x,x')\beta^n~,   
\end{align}
where the $a_n$ are the so-called Seleey-DeWitt coefficients and $\sigma(x,x')$ is the Synge function which amounts to half the geodesic distance between $x$ and $x'$.

In the path integral approach to the computation of the heat kernel in curved space a few issues have to be tackled. Firstly, space-time reparametrization invariance must be implemented both in the action and in the path integral measure, i.e. formally such measure reads
\begin{align}
    {\cal D}x = \prod_{0<t<\beta}\prod_{\mu} \sqrt{g(x(t))} dx^\mu(t) ~,
\end{align}
which, in turn, can be exponentiated by means of Lee-Yang ghosts (first introduced in this context by Bastianelli in Ref.~\cite{Bastianelli:1991be})
\begin{align}
  {\cal D}x = Dx \int Da Db Dc~e^{-\int_0^\beta dt \, \frac12 g_{\mu\nu} (x(t))\big( a^\mu a^\nu(t) + b^\mu c^\nu(t)\big)}~,   \end{align}
where, with $D\cdot$, we indicate Poincar\'e-invariant measures, i.e.
\begin{align}
 Dx =  \prod_{0<t<\beta}\prod_{\mu} dx^\mu(t)~,  
\end{align}
for the coordinates, and similarly for the ``commuting ghosts'' $a^\mu$, and for the  ``anticommuting ghosts'' $b^\mu$ and $c^\mu$. Moreover, a specific regularization scheme must be implemented in order to correctly compute the path integral, and suitable counterterms must be provided. In the present manuscript we employ ``Dimensional Regularization'' in the short time perturbative approach to the transition amplitude. 

Thus, the transition amplitude can be written as
\begin{align}
    K(x,x';\beta) = \int_L Dx Da Db Dc~e^{-\int_{-1}^0 d\tau \, \big[\frac1{2\beta} g_{\mu\nu}(x)\big( \dot x^\mu \dot x^\nu +a^\mu a^\nu +b^\mu c^\nu\big) +\beta(V(x)+V_{DR}(x))\big]}
    \label{eq:TA}
\end{align}
where we have used a rescaled time $\tau =\tfrac{t}{\beta}-1$ (and also redefined the ghost fields to absorb a factor of $\beta$), $V_{DR}$ is the aforementioned counterterm potential and $L$ refers to the fact that we are integrating over a ``line'' topology, i.e. with paths satisfying the boundary conditions
\begin{align}
    L:\ x^\mu(-1) =x^\mu\,,\quad x^\mu(0) =x'^\mu
\end{align}
whereas ghost fields have vanishing Dirichlet boundary conditions. 

In such a case, it is convenient to split the coordinate path as the sum of a background which satisfies the flat space geodesic equation, and quantum fluctuations
\begin{align}
    x^\mu(\tau) = \underbrace{x'^\mu -\xi^\mu \tau}_{x^\mu_{bg}(\tau)} +q^\mu(\tau)\,,\quad \xi^\mu:= x^\mu -x'^\mu\,,\quad q^\mu(0)=q^\mu(-1)=0\,,
    \label{eq:back-q}
\end{align}
and Taylor expand the metric about the final point. In particular, we will be using a particular set of coordinates, i.e. Riemann Normal Coordinates (RNC) centered at the point $x'^\mu$, taken as the origin, i.e. $x'^\mu|_{\rm RNC}=0$. The metric expansion (to the needed order) thus reads
\begin{align}
    g_{\mu\nu} (z) &=\underbrace{g_{\mu\nu} (0)}_{\delta_{\mu\nu}}+\frac13 R_{\alpha\mu\nu\beta}(0) z^\alpha z^\beta  +\frac16 \nabla_\gamma R_{\alpha\mu\nu\beta}(0) z^\alpha z^\beta z^\gamma\nonumber\\
    &+\Big[\frac1{20} \nabla_\delta\nabla_\gamma R_{\alpha\mu\nu\beta}(0)+\frac{2}{45} R_{\alpha\mu\sigma\beta}R_{\gamma\nu{}^\sigma\delta}(0)\Big]z^\alpha z^\beta z^\gamma z^\delta +O(z^5)\,,
\end{align}
where $z^\mu$ is a generic point on the manifold, in RNC, and the underbrace in the previous expression is meant to convey the information that we can always rotate the metric at the origin to the flat metric. Note that the choice of RNC is particularly convenient, since the linear term is absent in the metric expansion. 

The action in Eq.~\eqref{eq:TA}, splits into a quadratic part with flat metric for the quantum fields (the linear parts in $\dot q$ integrates to zero), 
\begin{align}
    S_{2}[\xi;q,a,b,c] = \frac{1}{2\beta} \delta_{\mu\nu}\xi^\mu \xi^\nu +\frac{1}{2\beta} \int_{-1}^0 d\tau\,  \delta_{\mu\nu} \big(\dot q^\mu \dot q^\nu +a^\mu a^\nu +b^\mu c^\nu \big)
    \label{eq:S2}
\end{align}
and an interaction part
\begin{align}
   S_{int}[\xi;q,a,b,c] = S_4+S_5+S_6+\cdots~, 
\end{align}
where,
\begin{align}
   S_4 =&\int_{-1}^0d\tau\,\tfrac1{6\beta} R_{\alpha\mu\nu\beta}(0) (-\xi^\alpha \tau +q^\alpha)(-\xi^\beta \tau + q^\beta) \Big[ \big(-\xi^\mu +\dot q^\mu) \big(-\xi^\nu +\dot q^\nu) +a^\mu a^\nu +b^\mu c^\nu\Big]+\nonumber\\ &+\beta \tilde V(0)~,\\
   S_5 =& \int_{-1}^0d\tau\,\tfrac1{12\beta} \nabla_\gamma R_{\alpha\mu\nu\beta}(0) (-\xi^\alpha \tau +q^\alpha)(-\xi^\beta \tau + q^\beta) (-\xi^\gamma \tau +q^\gamma) \Big[ \big(-\xi^\mu +\dot q^\mu) \big(-\xi^\nu +\dot q^\nu)\nonumber\\& +a^\mu a^\nu +b^\mu c^\nu\Big]+\beta \int_{-1}^0d\tau\,\partial_\alpha \tilde V(0) (-\xi^\alpha \tau +q^\alpha)~,\\
   S_6 = & \int_{-1}^0d\tau\, \Big(\tfrac1{40\beta} \nabla_\delta\nabla_\gamma R_{\alpha\mu\nu\beta}(0)+\tfrac{1}{45\beta} R_{\alpha\mu\sigma\beta}R_{\gamma\nu}{}^\sigma{}_\delta(0)\Big)(-\xi^\alpha \tau +q^\alpha)(-\xi^\beta \tau + q^\beta) (-\xi^\gamma \tau +q^\gamma)\nonumber\\
   &(-\xi^\delta \tau + q^\delta)\Big[ \big(-\xi^\mu +\dot q^\mu) \big(-\xi^\nu +\dot q^\nu) +a^\mu a^\nu +b^\mu c^\nu\Big]\nonumber\\
   &+\int_{-1}^0d\tau\,\tfrac\beta 2 \partial_\alpha \partial_\beta \tilde V(0)(-\xi^\alpha \tau +q^\alpha) (-\xi^\beta \tau + q^\beta)
\end{align}
and we have defined $\tilde V:= V +V_{DR}$. The expansion has been displayed only till the order needed below and the underscripts refer to the number of legs presents in the different vertices---note that potential terms are of order $\beta^2$ higher than the metric expansion, thus they effectively have four more legs than the corresponding metric expansion terms.  Indeed, as far as the $\beta$ power counting goes, each leg (background $\xi^\alpha$ or quantum fluctuation $q^\alpha$) yields a power $\tfrac12$, since the background has a Gaussian weight $\xi^2/2\beta$---see Eq.~\eqref{eq:S2}---, whereas the fluctuations have Green's functions which are linear in $\beta$---see Eq.~\eqref{eq:Greens} below. Each vertex, which comes from the expansion of the metric, yields a power $-1$, whereas vertices that are associated to the power-series expansion of the potential are worth it a power $+1$.

From~\eqref{eq:S2} one obtains the Green's functions (propagators) for the quantum fields, which can be written as
\begin{equation}
\begin{split}
    & \langle q^\mu(\tau) q^\nu(\sigma)\rangle = -\beta \delta^{\mu\nu} \Delta(\tau,\sigma)\\
    & \langle a^\mu(\tau) a^\nu(\sigma)\rangle = \beta \delta^{\mu\nu} \Delta_{gh}(\tau,\sigma)\\
    & \langle b^\mu(\tau) c^\nu(\sigma)\rangle = -2\beta \delta^{\mu\nu}\Delta_{gh}(\tau,\sigma)
   \end{split}
    \label{eq:Greens} 
\end{equation}
where
\begin{align}
    {}^{\bullet\bullet}\Delta(\tau,\sigma) = \Delta_{gh}(\tau,\sigma) =\delta(\tau,\sigma)~,  
\end{align}
where left (right) dot means derivative with respect to $\tau$ ($\sigma$)---the functional form of the $q q$ propagator and its derivatives are reported in the Appendix~\ref{sec:App-A}. Here it is important to stress the well-known feature of the interacting action above, that it presents derivative interactions, which give rise to superficially divergent Feynman diagrams and ambiguities, that are associated to products of distributions. For this simple reason, which has been known for decades it is necessary to regulate the computation of single Feynman diagrams which contribute to the short-time expansion of the transition amplitude. In fact, although the importance of regularization has been known for long time---and it sometimes lead to erroneous results---, it seems to have been neglected even in recent publications, which make use of such transition amplitudes in curved spaces.

\section{The Short-time Expansion of the Transition Amplitude}
\label{sec:TTA}
The expression of the transition amplitude at order $\beta^2$, is given by the three-loop expansion of the worldline action described in the previous section, i.e. by using the (Gaussian) path integral average of the interacting action.  This can be easily seen by using the power counting rules described above. For instance, the term $S_6$ provides a contribution of order $\beta^2$. Its first addendum, comes from the metric expansion: it has an overall $\tfrac1\beta$ factor, and has six coordinates, which count as $\beta^3$, and gives rise to a three-loop `daisy' diagram. 

Thus, given the free path integral normalization
\begin{align}
 \int_{q(-1)=0}^{q(0)=0} Dq Da Db Dc~e^{-\int_{-1}^0 d\tau \, \frac1{2\beta} \delta_{\mu\nu}\big( \dot q^\mu \dot q^\nu +a^\mu a^\nu +b^\mu c^\nu\big)}   = \frac{1}{(2\pi\beta)^{D/2}}~,
\end{align}
we obtain
\begin{align}
    K(x,x';\beta) &= \frac{e^{-\tfrac1{2\beta} \delta_{\mu\nu} \xi^\mu \xi^\nu}}{(2\pi\beta)^{D/2}}\, \Big\langle e^{-S_{int}}\Big\rangle\nonumber\\
    &= \frac{e^{-\tfrac1{2\beta} \delta_{\mu\nu} \xi^\mu \xi^\nu}}{(2\pi\beta)^{D/2}}\, e^{-\big(\big\langle S_4\big\rangle+\big\langle S_5\big\rangle + \big\langle S_6\big\rangle\big)+\tfrac12 \big\langle S_4^2\big\rangle_{conn}+\cdots\big)}\nonumber\\
    &= \frac{e^{-\tfrac1{2\beta} \delta_{\mu\nu} \xi^\mu \xi^\nu}}{(2\pi\beta)^{D/2}}\,\Big[1 - \big(\big\langle S_4\big\rangle+\big\langle S_5\big\rangle + \big\langle S_6\big\rangle\big)+\tfrac12 \big\langle S_4^2\big\rangle_{conn} +\tfrac12 \big\langle S_4\big\rangle^2+\cdots\Big]~,
    \label{eq:TA-fin}
\end{align}
and the expectation values are computed using Wick's theorem with the Green's functions~\eqref{eq:Greens}, and derivatives thereof. They read (all the tensors are evaluated at the origin of RNC),
\begin{align}
  \big\langle S_4\big\rangle =&  -\tfrac16 R_{\alpha\beta}\, \xi^\alpha \xi^\beta\, I_1 +\tfrac{\beta}{6}R\, I_2+\beta \tilde V\\
  \big\langle S_5\big\rangle =& \tfrac{1}{12}\nabla_\gamma R_{\alpha \beta}\, \xi^\alpha \xi^\beta \xi^\gamma\, I_3-\tfrac{\beta}{6}\nabla_\alpha R\, \xi^\alpha\, I_4+\tfrac{\beta}{2}\nabla_\alpha \tilde V\, \xi^\alpha \\
  \big\langle S_6\big\rangle =&-\big(\tfrac1{40} \nabla_\delta\nabla_\gamma R_{\alpha\beta}+\tfrac{1}{45} R_{\alpha\lambda\sigma\beta}R_{\gamma}{}^{\lambda\sigma}{}_\delta\big)\xi^\alpha \xi^\beta \xi^\gamma \xi^\delta\, I_5\nonumber\\& +\big(\tfrac{\beta}{40}\nabla^2 R_{\alpha \beta}+\tfrac{\beta}{20}\nabla^\lambda \nabla^\sigma R_{\alpha\lambda\sigma\beta} +\tfrac{\beta}{45} R_{\alpha\lambda} R_\beta{}^\lambda+\tfrac{\beta}{30}R_{\alpha\lambda\sigma\rho} R_\beta{}^{\lambda\sigma\rho}\big) \xi^\alpha \xi^\beta\, I_6\nonumber\\
  &+\big(\tfrac{\beta}{20}\nabla_\alpha\nabla_\beta R+\tfrac{2\beta}{45}R_{\alpha\lambda\sigma\beta} R^{\lambda\sigma}+\tfrac{\beta}{20}\nabla^\lambda\nabla_\alpha R_{\beta\lambda}-\tfrac{\beta}{20}\nabla^\lambda \nabla^\sigma R_{\alpha\lambda\sigma\beta} -\tfrac{\beta}{45} R_{\alpha\lambda} R_\beta{}^\lambda\nonumber\\& \quad\quad+\tfrac{\beta}{30}R_{\alpha\lambda\sigma\rho} R_\beta{}^{\lambda\sigma\rho}\big) \xi^\alpha \xi^\beta\, I_7\nonumber\\
  &
  -\big(\tfrac{\beta^2}{20} \nabla^2 R +\tfrac{\beta^2}{45} R_{\lambda\sigma}^2+\tfrac{\beta^2}{30} R_{\lambda\sigma\rho\tau}^2 \big)\, I_8
  -\tfrac{\beta^2}{2} \nabla^2 \tilde V\, I_9+\tfrac{\beta}{6}\nabla_\alpha\nabla_\beta \tilde V\, \xi^\alpha\xi^\beta\\
 \big\langle S_4^2\big\rangle_{conn} =&\tfrac{1}{36} R_{\alpha\lambda\sigma\beta} R_\gamma{}^{\lambda\sigma}{}_\delta\, \xi^\alpha \xi^\beta \xi^\gamma \xi^\delta\, I_{10}-\tfrac{\beta}{9}R_{\alpha\lambda}R_\beta{}^\lambda\, \xi^\alpha \xi^\beta\, I_{11}-\tfrac{\beta}{9}R_{\alpha\lambda\sigma \beta}R^{\lambda\sigma}\, \xi^\alpha \xi^\beta\, I_{12}\nonumber\\&
 -\tfrac{\beta}{6} R_{\alpha\lambda\rho\sigma} R_\beta{}^{\lambda\sigma\rho}\, \xi^\alpha \xi^\beta \, I_{13}+\tfrac{\beta^2}{18}R_{\lambda\sigma}^2\, I_{14}+ \tfrac{\beta^2}{12}R_{\lambda\sigma\rho\xi}^2\, I_{15}
\end{align}
where the worldline integrals $I_i\,,\ i=1,\dots, 15$, are given in the Appendix~\ref{sec:AppB}, both in the naive, unregulated, form and in the regulated form which is needed in order to compute them unambiguously, following the recipe given in the forthcoming Section~\ref{sec:DRTA}. To facilitate the comparison with earlier results, for the integrals we use the same conventions as in Ref.~\cite{Bastianelli:1998jb}.

\subsection{Dimensional Regularization of the Transition Amplitude}
\label{sec:DRTA}
The presence of derivative interactions make some of the Feynman diagrams, which contribute to the correlators listed above, superficially divergent and {\it a priori} ambiguous. Thus, we need to provide a regularization scheme to compute them. In the present manuscript we use and extend the version of Dimensional Regularization introduced in~\cite{Bastianelli:2000nm}. The main recipe is the dimensional extension of the compact worldline direction $-1\leq \tau=:t^0\leq 0$, with the addition of $D$ extra infinite dimensions, ${\mathbf{t}} =(t^1,\dots, t^D)$, i.e. the generic worldline integral gets generalized as
\begin{align}
  \int_{-1}^0d\tau   \quad \to \quad \int d^{D+1}t~.
\end{align}
The kinetic action for the $q^\mu$ fluctuations thus becomes
\begin{align}
    S_2[q] = \frac{1}{2\beta}\int d^{D+1}t\, \delta_{\mu\nu}\, \partial_a q^\mu \partial_a q^\nu 
\end{align}
where $\partial_a q^\mu := \tfrac{\partial}{\partial t^a} q^\mu$, and $t^a =(\tau, {\mathbf{t}})$.
Hence, the regularized Green's function for the fluctuations $q^\mu$ 
satisfies the $(1+D)$-dimensional Green's equation, 
\begin{align}
    \partial_a \partial_a \Delta(t,s)=:{}_{aa}\Delta(t,s)=\Delta_{gh}(t,s)= \delta(\tau,\sigma) \delta^{(D)}({\mathbf{t}}-{\mathbf{s}})
    \label{eq:GreensEq}
\end{align}
and, in the limit $D\to 0$, reduces to the unregulated Green's function---see Appendix~\ref{sec:App-A} for details on the Green's functions and conventions. Along with the previous rule, on the line we must also deal with a non-trivial background path. Since the extra dimensions are infinite and translation invariant, the background is taken to be non-trivial only on the compact direction, where we keep using the straight path described above in Eq.\eqref{eq:back-q}. Thus, in the regulated worldline, we have the following split between background and quantum fluctuations,
\begin{equation}
    \begin{split}
      & x^\mu (t) = -\xi^\mu \tau +q^\mu(t)\\
    & \partial_a x^\mu (t) = -\xi^\mu \delta_a^0 +\partial_a q^\mu(t)~.  
    \end{split}
    \label{eq:background}
\end{equation}
An important feature of the worldline version of DR is the use of integration by parts (i.b.p.'s). While on the extra dimensions i.b.p.'s is guaranteed (without boundary terms), thanks to Poincar\'e invariance, on the compact dimension more care is needed. On the one hand, the propagator $\Delta(t,s)$ vanishes at the end points of the compact direction, on both variables $\tau$ and $\sigma$, so the presence of underived propagators inside the integral ensures that no boundary terms are picked up upon i.b.p.'s. However, the derivatives of the propagator ${}^\bullet\Delta(t,s)$ and $\Delta^\bullet (t,s)$ only vanish at the end points, on the variable not involved in the derivative, i.e. $\sigma$ and $\tau$ respectively. When the two variables coincide, $t=s$, the regulated time derivative vanishes at the end points, whereas the unregulated time derivative is discontinuous there, as we point out in Appendix~\ref{sec:App-A}. One last, yet important, rule is the use of the Green's equation~\eqref{eq:GreensEq} at the regulated level, whose legitimacy was originally shown in Ref.~\cite{Bastianelli:2000nm}.

With this bag of tricks it is thus possible to unambiguously compute any worldline integral generated in the perturbative expansion of the nonlinear sigma model involved in the transition amplitude~\eqref{eq:TA}, i.e. for the wordline path integral on the line, in the same way as it was earlier done for path integrals on the circle, which are linked to one-loop effective actions in curved space.

Below, in order to clarify some details of how DR explicitely works for the computation of the Transition Amplitude, we give a few paradigmatic examples. 

Let us start with the integral 
\begin{align}
 I_1&=\int_{-1}^0d\tau\left[ \tau^2\left( {}^{\bullet}\Delta^{\bullet}+{}^{\bullet\bullet} \Delta\right)+\Delta-2\tau\, {}^{\bullet}\Delta \right]\vert_{\tau}   
\end{align}
which involves equal time contractions, and double derivatives and ghosts which separately yield divergent contributions. The worldline dimensional extension provides the prescription
\begin{align}
 I_1&  \to\int dt\left[ \tau^2\left( {}_a\Delta_a+{}_{aa}\Delta \right)+\Delta-2\tau\, {}_a\Delta \right]\vert_{t}~.  
\end{align}
The second and third term are unambiguous and can be computed directly at the unregulated level using the Green's functions~\eqref{eq:dot|}. In order to treat the first addendum one uses the identity $( {}_a \Delta_a +  {}_{aa}\Delta)|_t  ={}_0 ( {}_0 \Delta|_t)= {}^\bullet ( {}^\bullet \Delta|_t)$ discussed in the Appendix~\ref{sec:App-A}, and originally derived in Ref.~\cite{Bastianelli:2000nm}. Thus, by performing an integration by parts, which is allowed because, at the regulated level, ${}^\bullet \Delta|_t$ vanishes at both endpoints, one gets
\begin{align}
  I_1&  \to  \int dt\left[\Delta-4\tau\, {}_a\Delta \right]\vert_{t} \ \to \ \int_{-1}^0d\tau\left[\Delta-4\tau\,{}^{\bullet}\Delta \right]\vert_{\tau}  = -\frac12 ~. 
\end{align}
This rule is sufficient to unambiguously computed all the single integrals $I_i,\ i=1,\dots,9$, whose numerical result is reported in Appendix~\ref{sec:AppB}. 

Double integrals, which appear in the contraction $\langle S_4^2\rangle_{conn}$, are obviously more involved, though mostly straightforward. In order to better clarify how DR works, let us point out a few {\it a priori} tricky passages, by detailing the computation of the integrals $I_{10}$ and $I_{14}$. For the former, we have
\begin{align}
    I_{10}&=\int_{-1}^0d\tau\int_{-1}^0d\sigma\Big[2\tau^2\left( {}^{\bullet}\Delta^{\bullet}\;{}^{\bullet}\Delta^{\bullet}-{}^{\bullet\bullet}\Delta\Delta^{\bullet\bullet} \right)\sigma^2+4\tau^2({}^{\bullet}\Delta)^2+\nonumber\\
        &-8\tau^2{}^{\bullet}\Delta\;{}^{\bullet}\Delta^{\bullet}\sigma+2\Delta^2-8\Delta\Delta^{\bullet}\sigma+4\tau\Delta\;{}^{\bullet}\Delta^{\bullet}\sigma+4\tau{}^{\bullet}\Delta\Delta^{\bullet}\sigma  \Big]~.
\end{align}
Again, the ambiguous parts are those which involve double derivatives on the same Green's functions, i.e. the first, third and sixth terms. At the dimensionally extended level it reads
\begin{align}
  I_{10} &\to\int dt\int ds\Big[2\tau^2\left( {}_a\Delta_b\;{}_a\Delta_b-{}_{aa}\Delta\Delta_{bb} \right)\sigma^2+4\tau^2({}_a\Delta)^2+\nonumber\\
    &-8\tau^2{}_a\Delta\;{}_a\Delta_0\, \sigma+2\Delta^2-8\Delta\Delta_0\, \sigma+4\tau\Delta\;{}_0\Delta_0\, \sigma+4\tau\, {}_0\Delta\Delta_0\, \sigma \Big]
    \label{eq:I10-DR}
    \end{align}
    Firstly, note that there are terms where the derivative does not get dimensionally extended; the reason is that in those cases there is a contraction with a background field velocity, which is $\partial_a x^\mu(t)|_{bg} = -\delta_a^0 \xi^\mu$---see eq.~\eqref{eq:background}. Thus, for instance, $\Delta_a \partial_a x^\mu|_{bg} =-\Delta_0\, \xi^\mu$ and only the derivative with respect the compact direction remains on the Green's function. On the first term we can integrate by parts the $t^a$ and $t^b$ derivatives of the first addendum till we reach an expression which cancels the ghost contribution. The leftover reads
    \begin{align}
        \int dt\int ds\Big[ -4 \tau^2{}_a\Delta\;{}_a\Delta_0\, \sigma +4 \tau\, {}_0\Delta \Delta_{bb}\, \sigma^2\Big]~,
    \end{align}
    where the first contribution just renormalizes the coefficient of the third term of~\eqref{eq:I10-DR}, whereas the second contribution can be easily computed, since at regulated DR level one can safely use the Green's equation, as shown in~\cite{Bastianelli:2000nm}, and then remove the regularization. Thus,
    \begin{align}
        \int dt\int ds\, 4 \tau\, {}_0\Delta \Delta_{bb}\, \sigma^2 =  4 \int_{-1}^0 d\tau\, \tau^3\, {}^\bullet \Delta|_{\tau} = \frac{3}{10}~. 
    \end{align}
    The third term involves the expression $\tau^2 {}_a\Delta\;{}_a\Delta_0\, \sigma =\tau^2 \tfrac12 ({}_a\Delta^2)_0\, \sigma$, which can be integrated by parts on the right variable, since ${}_a\Delta^2$ vanishes at both endpoints on the right variable: it provides a term equal (apart for the coefficient) to the second one. Similarly, the sixth term can be integrated by parts with respect to, say, the left variable, and provides a contribution equal to (minus) the last term and one which is half of the fifth one. 
    \begin{align}
        I_{10}\ & \to\  \frac{3}{10} + \int dt\int ds\Big[4\tau^2({}_a\Delta)^2-12\tau^2{}_a\Delta\;{}_a\Delta_0\, \sigma
        +2\Delta^2 -12 \Delta \Delta_0\, \sigma\Big]\nonumber\\
        &  \to\ \frac{3}{10} + \int dt\int ds\Big[10\tau^2({}_a\Delta)^2
        +8\Delta^2\Big]\ \to\ \frac{3}{10} + \int_{-1}^0 d\tau\int_{-1}^0 d\sigma\Big[10\tau^2\, ({}^\bullet \Delta)^2 +8\Delta^2\Big] 
        =1~,
    \end{align}
    where we have used again integration by parts to get rid of the double derivative and then have removed the regularization in the final unambiguous form.
    
    Let us now focus on the integral $I_{14}$ which presents a little subtlety, that is worth it to discuss. The first term in $I_{14}$, which involves ghost contributions, can be dealt with as in the previous example. The second term reads
    \begin{align}
     I_{14,2}:= \int_{-1}^0d\tau \int_{-1}^0 d\sigma \Delta|_\tau {}^\bullet\Delta^\bullet\ {}^\bullet\Delta\ {}^\bullet\Delta|_\sigma\ \to\  \int dt \int ds\, \Delta|_t\ {}_a\Delta_0\ {}_a\Delta\ {}_0\Delta|_s~.
      \end{align}
Thus, by integration by parts we can remove the second derivative in the second Green's function. However, note that this would produce a second derivative on the last Green's function, i.e.
\begin{align}
I_{14,2}\ \to\  -\tfrac12  \int dt \int ds\, \Delta|_t\ ({}_a\Delta)^2\ {}_0({}_0\Delta|_s)
\end{align}
which must be held with care. In fact, at the unregulated level, the expression ${}^\bullet\Delta|_\tau$ gets periodically extended beyond the interval $(-1,0)$, but is discontinuous at both end points. Thus, its derivative involves delta functions at the endpoints, and could not {\it a priori} be set to unity. However, it is integrated with a smooth function which vanishes at both endpoints, and one can thus safely set it to unity---see Appendix~\ref{sec:App-A} for a more detailed explanations. Hence,
\begin{align}
    I_{14,2} = -\tfrac12 \int_{-1}^0d\tau \int_{-1}^0 d\sigma \Delta|_\tau ({}^\bullet \Delta)^2 =\frac{1}{90}~,
\end{align}
where we have used the expressions given in eq.~\eqref{eq:dot|}. In turn, the above rules implies that, for example, the last term of $I_{14}$ can be quite easily computed, i.e.
\begin{align}
    I_{14,7}&:=\int_{-1}^0d\tau \int_{-1}^0 d\sigma\, ({}^{\bullet}\Delta^{\bullet}+{}^{\bullet\bullet}\Delta)\vert_{\tau}({}^{\bullet}\Delta^{\bullet}+{}^{\bullet\bullet}\Delta)\vert_{\sigma}\Delta^2\nonumber\\
    &\to\ \int dt\int ds\ {}_0({}_0\Delta|_t)\ {}_0({}_0\Delta|_s)\ \Delta^2 \ \to\  
    \int_{-1}^0d\tau \int_{-1}^0 d\sigma \, \Delta^2 =\frac{1}{90}\ .
\end{align}
All other terms can be treated in a similar way.

Let us conclude with a few comments on the other double integrals, which can be straightforwardly computed with the rules described above. Of those, $I_{13}$ and $I_{15}$---although they are less subtle to compute than the ones described before, due to the more straightforward application of integration by parts---are the only ones which give different results as compared to other regularization schemes. In fact, they are the only ones which are genuinely, at least, two loops, i.e. they do not involve equal time loops as all the others. 

Thus, with the results listed in the Appendix~\ref{sec:AppB} for the integrals $I_i$, one can immediately check that the expression given in Eq.~\ref{eq:TA-fin} correctly reproduces the three-loop transition amplitude (see e.g.~\cite{Bastianelli:1998jb}), provided the counterterm potential is fixed to be
\begin{align}
    V_{DR} = \frac18 R~. 
\end{align}


\section{Conclusions and Outlook}
\label{sec:concl}
In the present manuscript we have extended the application of worldline Dimensional Regularization to path integrals in curved spaces defined on the line, i.e. with open boundary conditions, as opposed to previous constructions, which focused on path integrals with periodic boundary conditions. Specific rules of application of such regularization method are explained, using a three-loop computation of the particle transition amplitude.

The simplicity of the counterterm potential associated to Dimensional Regularization---which, unlike other schemes, only involves  covariant terms---may help in the use of such formalism to obtain open line (and one loop) master formulas with an arbitrary number of gravitons, though the need of integration by parts to remove divergences may somehow complicate the straightforwardness of the method. Moreover, in general, in the worldline computation of scattering amplitudes involving gravitons, one of the main complications comes from fact that the particle coordinates are coupled to the curved metric in a generic (non linear) way. As mentioned above in Section~\ref{sec:DR}, the scalar potential may involve couplings to the scalar curvature, which is thus renormalized by the regularization counterterm, i.e. $\tilde V =(\alpha+\tfrac18)R$. Therefore, graviton vertex operators become increasingly complicated as the number of gravitons grows. However, an exceptionally simple case occurs when the scalar potential is cancelled by the DR counterterm (i.e. $\alpha =-\tfrac18$), giving rise to a particle action where the coupling to gravity is linear, so only a single-graviton vertex operator is present. In this worldline-minimal scenario---which is the one adopted in the classical scattering computations of Ref.~\cite{Mogull:2020sak}---our DR regularization would provide a quite helpful simplification. However, as pointed out, for the scalar particle such coupling is a very special one. On the other hand, for a Dirac particle coupled to gravity, which  in the worldline approach, can be described by an ${\cal N}=1$ supersymmetric particle in curved space, the hamiltonian is computed from the square of the supersymmetry operator which produces a scalar curvature term that exactly cancels the DR counterterm and leaves a linear coupling to gravity. It would be thus desirable to extend our open-line construction to a spinning particle in curved space.

\paragraph{Acknowledgements}
The Authors thank Fiorenzo Bastianelli for helpful comments and suggestions.

\appendix

\section{Conventions and Green's functions}
\label{sec:App-A}
Throughout the manuscript we use the following conventions
\begin{align}
    &\big[ \nabla_\mu,\nabla_\nu\big] V^\sigma = R_{\mu\nu}{}^\sigma{}_\rho V^\rho\\
    & R_{\mu\nu} = R_{\mu\sigma}{}^\sigma{}_\rho
\end{align}
i.e. we take spheres with negative scalar curvature.

In the dimensionally extended worldline the Green's function for the coordinate fluctuations---which have Dirichlet boundary conditions on the compact direction---reads
\begin{align}
\big\langle q^\mu (t) q^\nu(s)\big\rangle=-\beta \delta^{\mu\nu}\Delta(t,s)
\end{align}
which satisfies the equation
\begin{align}
  {}_{aa}\Delta(t,s)= \delta(\tau,\sigma)\delta^{(D)}({\mathbf{t}}-{\mathbf{s}})~.  
\end{align}
Henceforth, we denote the derivative with respect to the left variable $\partial/\partial t^a$ with a left subscript $a$, and the derivative with respect to the right variable $\partial/\partial s^a$ with a right subscript $a$, i.e.
\begin{align}
    &\tfrac{\partial}{\partial t^a} \Delta(t,s) =: {}_a \Delta(t,s)\\
    & \tfrac{\partial}{\partial s^a} \Delta(t,s) =: \Delta_a(t,s)\\
    &\tfrac{\partial^2}{\partial t^a\partial s^b} \Delta(t,s) =: {}_a \Delta_b(t,s)~.
\end{align}
In momentum space the Green's function can be written as
\begin{align}
  \Delta(t,s) =\int \frac{d^Dk}{(2\pi)^D}\sum_{n=1}^\infty \frac{-2}{(\pi n)^2 +{\mathbf{k}}^D} \sin(\pi n\tau)\sin(\pi n\sigma) e^{i{\mathbf{k}}\cdot {\mathbf{(t-s)}}}~,  
  \label{eq:reg-green}
\end{align}
which implies the following helpful identity
\begin{align}
 {}_a \Delta|_t:={}_a \Delta(t,t) =\delta_{a0} {}^\bullet \Delta|_t
 \label{eq:id1}
\end{align}   
where left (right) bullet indicates derivative with respect to the left (right) compact time variable---note that the expression~\eqref{eq:id1} only depends upon $\tau=t^0$. Thus,
\begin{align}
( {}_a \Delta_a +  {}_{aa}\Delta)|_t  ={}_0 ( {}_0 \Delta|_t)= {}^\bullet ( {}^\bullet \Delta|_t)
\label{eq:id2}
\end{align}
as was discussed, in more detail, in Ref.~\cite{Bastianelli:2000nm}. Since the regularization procedure prescribes to manipulate the integral expressions at the regulated level till unambiguous expressions are reached, which can be computed safely going to the unregulated limit $D\to 0$, it is helpful to report
the functional form of the Green's functions in such limit, i.e.
\begin{equation}
\begin{split}
    & \Delta(\tau,\sigma) = \tau (\sigma+1) \vartheta(\tau - \sigma) + \sigma (\tau+1) \vartheta(\sigma-\tau)\,,\\
    & {}^\bullet \Delta(\tau,\sigma)=\sigma+ \vartheta(\tau - \sigma)\,,\quad 
    \Delta^\bullet(\tau,\sigma)=\tau+ \vartheta(\sigma-\tau)\,,\\
    & \Delta|_\tau =\tau(\tau+1)\,, \\
    & {}^\bullet \Delta|_\tau= \tau +\tfrac12~. 
\end{split}
\label{eq:dot|}
\end{equation}
From the latter it would be tempting to conclude that, 
\begin{align}
  {}^\bullet ( {}^\bullet \Delta|_t) \stackrel{{D\to 0}}{\longrightarrow} 1  ~.
  \label{eq:id3}
\end{align}
 However, here a few clarification are in order. The regulated Fourier series~\eqref{eq:reg-green} vanishes at the end point, and so does ${}^\bullet \Delta|_t$. At the unregulated level the sum of the series must be periodically extended outside the interval $(-1,0)$, thus ${}^\bullet \Delta|_\tau$ is discontinuous at $\tau =0,-1$. Therefore its derivative has a singular (delta function like) part, at the boundary points. On the other hand, if the left hand side of~\eqref{eq:id3} multiplies an expression which vanishes at the boundary points, then it can be safely taken to be equal to $1$, which is what we do in various integrals below.      

\section{List of regulated integrals}
\label{sec:AppB}
In this section we give the list of worldline integrals which appear in our three-loop calculation. We express them in the naive, unregulated form, and in the regulated form (the arrow indicates the passage between the two forms), and give their final numerical result which, using the rules described above, can be straightforwardly obtained. 
\begin{align}
        I_1&=\int_{-1}^0d\tau\left[ \tau^2\left( {}^{\bullet}\Delta^{\bullet}+{}^{\bullet\bullet} \Delta\right)+\Delta-2\tau\, {}^{\bullet}\Delta \right]\vert_{\tau}\to\nonumber\\
        &\to\int dt\left[ \tau^2\left( {}_a\Delta_a+{}_{aa}\Delta \right)+\Delta-2\tau\, {}_0\Delta \right]\vert_{t}=-\frac{1}{2}\\
        I_2&=\int_{-1}^0d\tau\left[ \Delta\left( {}^{\bullet}\Delta^{\bullet}+{}^{\bullet\bullet}\Delta \right)-\left({}^{\bullet}\Delta\right)^2 \right]\vert_{\tau}\to\nonumber\\
        &\to\int dt\left[ \Delta\left( {}_a\Delta_a+{}_{aa}\Delta \right)-\left({}_a\Delta\right)^2 \right]\vert_{t}=-\frac{1}{4}\\
        I_3&=\int_{-1}^0d\tau\left[ \tau^3\left( {}^{\bullet}\Delta^{\bullet}+{}^{\bullet\bullet}\Delta \right)+\tau\Delta-2\tau^2\,{}^{\bullet}\Delta \right]\vert_{\tau}\to\nonumber\\
        &\to\int dt\left[ \tau^3\left( {}_a\Delta_a+{}_{aa}\Delta \right) +\tau\Delta-2\tau^2\, {}_0\Delta\right]\vert_t=\frac{1}{2}\\
        I_4&=\int_{-1}^0d\tau\left[ \tau\Delta\left( {}^{\bullet}\Delta^{\bullet}+{}^{\bullet\bullet}\Delta \right)-\tau({}^{\bullet}\Delta)^2 \right]\vert_{\tau}\to\nonumber\\
        &\to\int dt\left[ \tau\Delta\left( {}_a\Delta_a+{}_{aa}\Delta \right)-\tau({}_a\Delta)^2 \right]\vert_t=\frac{1}{8}\\
        I_5&=\int_{-1}^0d\tau\left[ \tau^4\left( {}^{\bullet}\Delta^{\bullet}+{}^{\bullet
        \bullet}\Delta \right)+\tau^2\Delta-2\tau^3\, {}^{\bullet}\Delta \right]\vert_{\tau}\to\nonumber\\
        &\to\int dt\left[ \tau^4\left( {}_a\Delta_a+{}_{aa}\Delta \right)+\tau^2\Delta-2\tau^3\, {}_0\Delta \right]\vert_t=-\frac{1}{2}\\
        I_6&=\int_{-1}^0d\tau\left[ \tau^2\Delta\left( {}^{\bullet}\Delta^{\bullet}+{}^{\bullet\bullet}\Delta \right)+\Delta^2-2\tau\, {}^{\bullet}\Delta\Delta \right]\vert_{\tau}\to\nonumber\\
        &\to\int dt\left[ \tau^2\Delta\left( {}_a\Delta_a+{}_{aa}\Delta \right)+\Delta^2-2\tau\, {}_0\Delta\Delta \right]\vert_t=0\\
        I_7&=\int_{-1}^0d\tau\left[ \tau^2\Delta\left( {}^{\bullet}\Delta^{\bullet}+{}^{\bullet\bullet}\Delta \right)-\tau^2({}^{\bullet}\Delta)^2 \right]\vert_{\tau}\to\nonumber\\
        &\to\int dt\left[ \tau^2\Delta\left( {}_a\Delta_a+{}_{aa}\Delta \right)-\tau^2({}_a\Delta)^2 \right]\vert_t=-\frac{1}{12}\\
        I_8&=\int_{-1}^0d\tau\left[ \Delta^2\left( {}^{\bullet}\Delta^{\bullet}+{}^{\bullet\bullet}\Delta \right)-({}^{\bullet}\Delta)^2\Delta \right]\vert_{\tau}\to\nonumber\\
        &\to\int dt\left[ \Delta^2\left( {}_a\Delta_a+{}_{aa}\Delta \right)-({}_a\Delta)^2\Delta \right]\vert_t=\frac{1}{24}\\
        I_9&=\int_{-1}^0d\tau\Delta\vert_{\tau}=-\frac{1}{6}\\
        I_{10}&=\int_{-1}^0d\tau\int_{-1}^0d\sigma\Big[2\tau^2\left( {}^{\bullet}\Delta^{\bullet}\;{}^{\bullet}\Delta^{\bullet}-{}^{\bullet\bullet}\Delta\Delta^{\bullet\bullet} \right)\sigma^2+4\tau^2({}^{\bullet}\Delta)^2+\nonumber\\
        &-8\tau^2{}^{\bullet}\Delta\;{}^{\bullet}\Delta^{\bullet}\sigma+2\Delta^2-8\Delta\Delta^{\bullet}\sigma+4\tau\Delta\;{}^{\bullet}\Delta^{\bullet}\sigma+4\tau{}^{\bullet}\Delta\Delta^{\bullet}\sigma  \Big]\to\nonumber\\
        &\to\int dt\int ds\Big[2\tau^2\left( {}_a\Delta_b\;{}_a\Delta_b-{}_{aa}\Delta\Delta_{bb} \right)\sigma^2+4\tau^2({}_a\Delta)^2+\nonumber\\
        &-8\tau^2{}_a\Delta\;{}_a\Delta_0\,\sigma+2\Delta^2-8\Delta\Delta_0\,\sigma+4\tau\Delta\;{}_0\Delta_0\, \sigma+4\tau{}_0\Delta\Delta_0\,\sigma  \Big]=1\\
        I_{11}&=\int_{-1}^0d\tau\int_{-1}^0d\sigma\Big[ \tau({}^{\bullet}\Delta^{\bullet}+{}^{\bullet\bullet}\Delta)\vert_{\tau}\Delta({}^{\bullet}\Delta^{\bullet}+{}^{\bullet\bullet}\Delta)\vert_{\sigma}\sigma+\tau{}^\bullet\Delta\vert_{\tau}\;{}^{\bullet}\Delta^{\bullet}\ {}^\bullet\Delta\vert_{\sigma}\sigma+\nonumber\\
        &-2\tau({}^{\bullet}\Delta^{\bullet}+{}^{\bullet\bullet}\Delta)\vert_{\tau}\Delta^{\bullet}\ {}^\bullet\Delta\vert_{\sigma}\sigma+\Delta\vert_{\tau}\;{}^{\bullet}\Delta^{\bullet}\;\Delta\vert_{\sigma}+{}^\bullet\Delta\vert_{\tau}\, \Delta\ {}^\bullet\Delta\vert_{\sigma}-2\Delta\vert_{\tau}\;{}^{\bullet}\Delta\ {}^\bullet\Delta\vert_{\sigma}+\nonumber\\
        &+2\tau({}^{\bullet}\Delta^{\bullet}+{}^{\bullet\bullet}\Delta)\vert_{\tau}\Delta^{\bullet}\;\Delta\vert_{\sigma}+2\tau\, {}^\bullet\Delta\vert_{\tau}\ {}^{\bullet}\Delta\ {}^\bullet\Delta\vert_{\sigma}+\nonumber\\
        &-2\tau({}^{\bullet}\Delta^{\bullet}+{}^{\bullet\bullet}\Delta)\vert_{\tau}\Delta\ {}^\bullet\Delta\vert_{\sigma}-2\tau\, {}^\bullet\Delta\vert_{\tau}\;{}^{\bullet}\Delta^{\bullet}\;\Delta\vert_{\sigma} \Big]\to\nonumber\\
        &\to\int dt\int ds\Big[ \tau({}_a\Delta_a+{}_{aa}\Delta)\vert_t\,\Delta({}_b\Delta_b+{}_{bb}\Delta)\vert_s\,\sigma+\tau\Delta_b\vert_t\;{}_a\Delta_b\;\Delta_b\vert_s\sigma+\nonumber\\
        &-2\tau({}_a\Delta_a+{}_{aa}\Delta)\vert_t\; \Delta_b\;\Delta_b\vert_s\,\sigma+\Delta\vert_t\;{}_0\Delta_0\;\Delta\vert_s+{}_0\Delta\vert_t\; \Delta\,{}_0\Delta\vert_s-2\Delta\vert_t\;{}_0\Delta\, {}_0\Delta\vert_s+\nonumber\\
        &+2\tau({}_a\Delta_a+{}_{aa}\Delta)\vert_t\; \Delta_0\;\Delta\vert_s+2\tau\, {}_a\Delta\vert_t\;{}_a\Delta\; {}_0\Delta\vert_s+\nonumber\\
        &-2\tau({}_a\Delta_a+{}_{aa}\Delta)\vert_t\; \Delta\; {}_0\Delta\vert_s-2\tau{}_a\Delta\vert_t\;{}_a\Delta_0\;\Delta\vert_s \Big]=-\frac{1}{12}\\
        I_{12}&=\int_{-1}^0d\tau\int_{-1}^0d\sigma\Big[ \tau^2\Delta\vert_{\sigma}\left( {}^{\bullet}\Delta^{\bullet}\;{}^{\bullet}\Delta^{\bullet}-{}^{\bullet\bullet}\Delta\Delta^{\bullet\bullet} \right)+(\Delta^{\bullet})^2\Delta\vert_{\sigma}+\nonumber\\
        &-2\tau\Delta^{\bullet}\;{}^{\bullet}\Delta^{\bullet}\;\Delta\vert_{\sigma}+\tau^2({}^{\bullet}\Delta)^2({}^{\bullet}\Delta^{\bullet}+{}^{\bullet\bullet}\Delta)\vert_{\sigma}+\Delta^2({}^{\bullet}\Delta^{\bullet}+{}^{\bullet\bullet}\Delta)\vert_{\sigma}-2\tau\Delta\;{}^{\bullet}\Delta({}^{\bullet}\Delta^{\bullet}+{}^{\bullet\bullet}\Delta)\vert_{\sigma} +\nonumber\\&-2\tau^2{}^{\bullet}\Delta\:{}^{\bullet}\Delta^{\bullet}\;\Delta^{\bullet}\vert_{\sigma}
        -2\Delta\Delta^{\bullet}\;\Delta^{\bullet}\vert_{\sigma}+2\tau\Delta\;{}^{\bullet}\Delta^{\bullet}\;\Delta^{\bullet}\vert_{\sigma}+2\tau{}^{\bullet}\Delta\Delta^{\bullet}\;\Delta^{\bullet}\vert_{\sigma} \Big]\to\nonumber\\
        &\to\int dt\int ds\Big[ \tau^2\Delta\vert_s\left( {}_a\Delta_b\;{}_a\Delta_b-{}_{aa}\Delta\Delta_{bb} \right)+(\Delta_b)^2\Delta\vert_s+\nonumber\\
        &-2\tau\Delta_b\;{}_a\Delta_b\;\Delta\vert_s+\tau^2({}_a\Delta)^2({}_a\Delta_b+{}_{aa}\Delta)\vert_s+\Delta^2({}_a\Delta_b+{}_{aa}\Delta)\vert_s-2\tau\Delta\;{}_a\Delta({}_a\Delta_b+{}_{aa}\Delta)\vert_s+\nonumber\\
        &-2\tau^2{}_a\Delta\;{}_a\Delta_b\;\Delta_b\vert_s-2\Delta\Delta_b\;\Delta_b\vert_s+2\tau\Delta\;{}_a\Delta_b\;\Delta_b\vert_s+2\tau{}_a\Delta\Delta_b\;\Delta_b\vert_s \Big]=\frac{1}{6}\\
        I_{13}&=\int_{-1}^0d\tau\int_{-1}^0d\sigma\Big[ \tau\sigma\Delta\left( {}^{\bullet}\Delta^{\bullet}\;{}^{\bullet}\Delta^{\bullet}-{}^{\bullet\bullet}\Delta\Delta^{\bullet\bullet} \right)-\tau\sigma{}^{\bullet}\Delta\Delta^{\bullet}\;{}^{\bullet}\Delta^{\bullet}+\nonumber\\
        &+\Delta^2\;{}^{\bullet}\Delta^{\bullet}-\Delta\;{}^{\bullet}\Delta\Delta^{\bullet}+2\tau\Delta^{\bullet}\;({}^{\bullet}\Delta)^2-2\tau\Delta\;{}^{\bullet}\Delta\;{}^{\bullet}\Delta^{\bullet} \Big]\to\nonumber\\
        &\to\int dt\int ds\Big[ \tau\sigma\Delta\left( {}_a\Delta_b\;{}_a\Delta_b-{}_{aa}\Delta\Delta_{bb} \right)-\tau\sigma{}_a\Delta\Delta_b\;{}_a\Delta_b+\nonumber\\
        &+\Delta^2\;{}_a\Delta_b-\Delta\;{}_a\Delta\Delta_b+2\tau\Delta_b\;({}_a\Delta)^2-2\tau\Delta\;{}_a\Delta\;{}_a\Delta_b \Big]=0\\
        I_{14}&=\int_{-1}^0d\tau\int_{-1}^0d\sigma\Big[ \Delta\vert_{\tau}\Delta\vert_{\sigma}\left( {}^{\bullet}\Delta^{\bullet}\;{}^{\bullet}\Delta^{\bullet}-{}^{\bullet\bullet}\Delta\Delta^{\bullet\bullet} \right)+\nonumber\\
        &-4\Delta\vert_{\tau}\;{}^{\bullet}\Delta^{\bullet}\;{}^{\bullet}\Delta\ {}^\bullet\Delta\vert_{\sigma}+2\Delta\vert_{\tau}({}^{\bullet}\Delta^{\bullet}+{}^{\bullet\bullet}\Delta)\vert_{\sigma}\;({}^{\bullet}\Delta)^2+2{}^\bullet\Delta\vert_{\tau}\Delta\;{}^{\bullet}\Delta^{\bullet}\ {}^\bullet\Delta\vert_{\sigma}+\nonumber\\
        &+2{}^\bullet \Delta\vert_{\tau}\;{}^{\bullet}\Delta\Delta^{\bullet}\ {}^\bullet \Delta\vert_{\sigma}-4{}^\bullet\Delta\vert_{\tau}({}^{\bullet}\Delta^{\bullet}+{}^{\bullet\bullet}\Delta)\vert_{\sigma}\Delta\;{}^{\bullet}\Delta+({}^{\bullet}\Delta^{\bullet}+{}^{\bullet\bullet}\Delta)\vert_{\tau}({}^{\bullet}\Delta^{\bullet}+{}^{\bullet\bullet}\Delta)\vert_{\sigma}\Delta^2 \Big]\to\nonumber\\
        &\to\int dt\int ds\Big[ \Delta\vert_t\Delta\vert_s\left( {}_a\Delta_b\;{}_a\Delta_b-{}_{aa}\Delta\Delta_{bb} \right)+\nonumber\\
        &-4\Delta\vert_t\;{}_{a}\Delta_b\;{}_a\Delta\Delta_b\vert_s+2\Delta\vert_t({}_b\Delta_b+{}_{bb}\Delta)\vert_s\;({}_a\Delta)^2+2\Delta_b\vert_t\Delta\;{}_a\Delta_b\;\Delta_b\vert_s+\nonumber\\
        &+2\Delta_b\vert_t\;{}_a\Delta\Delta_b\;\Delta_b\vert_s-4\Delta_a\vert_t\,({}_b\Delta_b+{}_{bb}\Delta)\vert_s\,\Delta\;{}_a\Delta+({}_a\Delta_a+{}_{aa}\Delta)\vert_t({}_b\Delta_b+{}_{bb}\Delta)\vert_s\Delta^2 \Big]=-\frac{1}{12}\\
        I_{15}&=\int_{-1}^0d\tau\int_{-1}^0d\sigma\Big[\Delta^2\left( {}^{\bullet}\Delta^{\bullet}\;{}^{\bullet}\Delta^{\bullet}-{}^{\bullet\bullet}\Delta\Delta^{\bullet\bullet} \right)+({}^{\bullet}\Delta)^2(\Delta^{\bullet})^2-2\Delta\;{}^{\bullet}\Delta\Delta^{\bullet}\;{}^{\bullet}\Delta^{\bullet} \Big]\to\nonumber\\
        &\to\int dt\int ds\Big[\Delta^2\left( {}_a\Delta_b\;{}_a\Delta_b-{}_{aa}\Delta\Delta_{bb} \right)+({}_a\Delta)^2(\Delta_b)^2-2\Delta\;{}_a\Delta\Delta_b\;{}_a\Delta_b \Big]=0.
\end{align}


\end{document}